\documentclass[journal]{IEEEtran}

\usepackage{cite}
\usepackage{color}
\usepackage{graphicx}
\usepackage[implicit=false]{hyperref}
\usepackage{epsfig}
\usepackage{amsmath, amssymb}

\usepackage{anyfontsize}
\usepackage{capt-of}
\usepackage{multirow}
\usepackage{array}
\usepackage{caption} 
\usepackage{flushend}
\usepackage[T1]{fontenc}
\usepackage{pslatex}
\usepackage{algorithm,algpseudocode}
\floatname{algorithm}{Algorithm}
\usepackage[table,xcdraw]{xcolor}
\usepackage{hhline}
\usepackage{tabularx}
\usepackage[nolist,,nohyperlinks]{acronym}
\usepackage{booktabs}

\begin{acronym}
\acro{ACK}{Acknowledgment}
\acro{ACN}{Availability Confirmation}
\acro{AODV}{ad hoc on-demand distance vector routing}
\acro{AP}{Access Point}
\acro{ARQ}{Automatic Reply Request}
\acro{ART}{Availability Request}
\acro{BE}{Backoff Exponent}
\acro{BER}{Bit Error Rate}
\acro{BI}{Beacon Interval}
\acro{CAP}{Contention Access Period}
\acro{CC}{Convolutional Coding}
\acro{CCA}{Clear Channel Assessment}
\acro{CC}{Control Channel}
\acro{CDMA}{Code Division Multiple Access}
\acro{CFP}{Contention Free Period}
\acro{CSMA/CA} {Carrier Sense Multiple Access/Collision Avoidance}
\acro{CSMA/CD} {Carrier Sense Multiple Access/Collision Detection}
\acro{CSK}{Color-Shift Keying}
\acro{CTS}{Clear-to-send}
\acro{DMT}{Discrete Multi-Tones}
\acro{DC}{Data Channel}
\acro{DCO-OFDM}{Direct-Current Optical \ac{OFDM}}
\acro{DCN}{Data Center Network}
\acro{D2D}{Device-to-Device}
\acro{DoA} {Direction of Arrival}
\acro{DoD}{Department Of Defence}
\acro{DSR}{dynamic source routing}
\acro{ECC}{Error-Correction Code}
\acro{EMI}{Electromagnetic Interference}
\acro{FEC}{Forward Error Correction}
\acro{FSO}{Free Space Optics}
\acro{FOV}{Field Of View}
\acro{GPS}{Global Positioning System}
\acro{GTS}{Guaranteed Time Slots}
\acro{IM/DD}{Intensity-Modulation Direct-Detection}
\acro{IoT}{Internet of Things}
\acro{IR} {Infra-Red}
\acro{ISM}{Industrial, Scientific and Medical}
\acro{ISR}{ Intelligence, Surveillance, and Reconnaissance}
\acro{LANET}{Visible Light Ad Hoc Network}
\acro{LED}{Light Emitting Diode}
\acro{Li-Fi}{Light Fidelity}
\acro{LOS}{Line of Sight}
\acro{LPI/LPD}{Lower Probability of Intercept/Lower Probability of Detection}
\acro{MAC}{Medium Access Control}
\acro{MA-DMT}{Multiple Access Discrete Multi-Tones}
\acro{MANET}{Mobile Ad Hoc Network}
\acro{MHC}{Minimum Hop Count }
\acro{MUI}{Multi-User Interference}
\acro{MU-MIMO}{Multi-User Multiple-Input Multiple-Output}
\acro{MU-MISO}{Multi-User Multiple-Input Single-Output}
\acro{NAV}{Network Allocation Vector}
\acro{NB}{Number of Backoffs}
\acro{MC-CDMA}{ Multi-carrier \ac{CDMA}}
\acro{NLOS}{non-Line Of Sight}
\acro{NRL}{Naval Research Labs}
\acro{OC}{Optical Carrier}
\acro{OCDMA}{Optical Code-Division Multiple Access}
\acro{OFDM}{Orthogonal Frequency Division Multiplexing}
\acro{OFDMA}{Orthogonal Frequency Division Multiple Access}
\acro{OOC}{Optical Orthogonal Codes}
\acro{O-OFDMA}{Optical Orthogonal Frequency Division Multiple Access}
\acro{O-OFDM-IDMA}{Optical Orthogonal Frequency Division Multiplexing Interleave Division Multiple Access}
\acro{OOK}{On-Off Keying}
\acro{OWC}{Optical Wireless Communication}
\acro{OWMAC}{Optical wireless MAC}
\acro{PD}{Photon Detector}
\acro{PHR}{PHY Header}
\acro{PHY}{Physical} 
\acro{PRO-OFDM}{Polarity Reversed Optical \ac{OFDM}}
\acro{PSDU}{PHY Service Data Unit}
\acro{QAM}{Quadrature Amplitude Modulation}
\acro{QoS}{Quality of Service}
\acro{RA}{Random Access}
\acro{RES}{Reserve Sectors}
\acro{RF}{Radio Frequency}
\acro{RLL}{Run Length Limited}
\acro{ROC}{Random Optical Codes}
\acro{RRS}{Route Reliability Score}
\acro{RS}{Reed-Solomon } 
\acro{SACW}{Self-Adaptive minimum Contention Window}
\acro{SD}{Superframe Duration}
\acro{S-IDLE}{Synchronous Idle State}
\acro{SNR}{Signal-to-Noise Ratio}
\acro{SWaP}{(Size, Weight, and Power)}
\acro{TDD}{Time Division Duplex}
\acro{TDMA}{Time Division Multiple Access}
\acro{TR}{Transceiving State}
\acro{VLC}{Visible Light Communication}
\acro{VLN}{Visible Light Node}
\acro{VPPM}{Variable Pulse Position Modulation }
\acro{VPAN}{Visible-light communication Personal Area Network}
\acro{USRP}{Universal Software Radio Peripheral}
\acro{UV}{Ultraviolet}
\acro{UVC}{Ultraviolet Communication}
\acro{VANET}{Vehicular Ad Hoc Networking}
\acro{V2I} {Vehicle to Infrastructure}
\acro{V2V} {Vehicle to Vehicle}
\acro{WiFi}{Wireless Fidelity}
\acro{WSN}{Wireless Sensor Network}
\acro{ZF}{Zero Forcing}
\acro{4B6B}{4-bit to 6-bit encoded symbols} 
\acro{8B10B}{8-bit to 10-bit encoded symbols}
\end{acronym}
\usepackage{stfloats}
\makeatletter
\newcommand{\thickhline}{%
    \noalign {\ifnum 0=`}\fi \hrule height 2pt
    \futurelet \reserved@a \@xhline
}
\newcolumntype{"}{@{\hskip\tabcolsep\vrule width 1pt\hskip\tabcolsep}}
\makeatother
\newcolumntype{?}{!{\vrule width 2pt}}

\begin{document}
\newcommand{\squeezeup}{\vspace{-8 mm}}
\newcommand{\squeezeuppp}{\vspace{-6 mm}}
\newcommand{\squeezeu}{\vspace{-5 mm}}
\newcommand{\squeezeupp}{\vspace{-3 mm}}
\newcommand{\squeezeupppp}{\vspace{-1 mm}}
\newcommand{\squeeze}{\vspace{-.3 mm}}
%
\title{VL-ROUTE: A Cross-Layer Routing Protocol for Visible Light Ad Hoc Network}

\author{\IEEEauthorblockN{Jithin Jagannath${^{\ddagger\dagger}}$, Tommaso Melodia${^\ddagger}$}

\IEEEauthorblockA{$^{\ddagger}$Northeastern University, Boston MA, \{jagannath.j, melodia\}@northeastern.edu}

\IEEEauthorblockA{$^{\dagger}$ANDRO Computational Solutions, LLC, Rome NY, jjagannath@androcs.com
}}

\maketitle

\begin{abstract}
Visible Light Ad Hoc Networks (LANETs) is being perceived as an emerging technology to complement Radio Frequency (RF) based ad hoc networks to reduce congestion in the overloaded RF spectrum. LANET is intended to support scenarios requiring dense deployment and high data rates. In Visible  Light  Communication  (VLC), most of the attention has been centered around physical layer with emphasis on point-to-point communication. In this work, we focus on designing a routing protocol specifically to overcome the unique challenges like blockage and deafness that render routes in LANETs highly unstable. Therefore, we propose a cross-layer optimized routing protocol (VL-ROUTE) that interacts closely with the Medium Access Control (MAC) layer to maximize the throughput of the network by taking into account the reliability of routes. 

To accomplish this in a distributed manner, we carefully formulate a Route Reliability Score (RRS) that can be computed by each node in the network using just the information gathered from its immediate neighbors. Each node computes an RRS for every known sink in the network. RRS of a given node can be considered as an estimate of the probability of reaching a given sink via that node. The RSS value is then integrated to the utility based three-way handshake process used by the MAC protocol (VL-MAC) to mitigate the effects of deafness, blockage, hidden node, and maximize the probability of establishing full-duplex links. All these factors contribute towards maximizing the network throughput. Extensive simulation of VL-ROUTE shows $\mathbf{124\%}$ improvement in network throughput over a network that uses Carrier Sense Multiple Access/Collision Avoidance (CSMA/CA) along with shortest path routing. Additionally, VL-ROUTE also showed up to $\mathbf{21\%}$ improvement in throughput over the network that uses VL-MAC along with a geographic routing. 
\end{abstract}
\begin{IEEEkeywords}
Visible light ad hoc network, cross-layer routing, 5G, reliability, visible light communication
\end{IEEEkeywords}

\IEEEpeerreviewmaketitle

\section{Introduction}

\ac{VLC} is envisioned as a major alternative to mitigate the congestion experienced by the current \ac{RF} spectrum. The recent emergence of \ac{VLC} along with the advancements in the enabling technologies such as \acp{LED} and \acp{PD} has lead to exploration of \acp{LANET} \cite{LANET}. Furthermore, \ac{VLC} has been argued to be a critical component of the 5th Generation(5G) technology providing data rates up to $8\;\mathrm{Gbps}$ \cite{Hass5G}. Accordingly, several indoor \cite{Coop-relay, Qin_datacenter} and outdoor applications \cite{VANET} have been identified in both commercial and military domains. 

\textbf{Indoor Applications.} Modern smart homes will have devices (TV, refrigerators thermostat among others) forming a \ac{LANET} and utilizing the ubiquitous lighting infrastructure to exchange large amounts of sensor data to provide enhance autonomy and efficiency. Employing \acp{LANET} for these applications will reduce the load on traditional networks like the \ac{WiFi} that are currently being used for these applications. In a commercial setting, to improve efficiency, \ac{VLC} is also envisioned as a medium of communication while constructing an inter-rack wireless \ac{DCN} \cite{Qin_datacenter}. These links are usually short and demand high data-rates which is where \acp{LANET} are most effective.

\textbf{Outdoor Civilian Applications.} \ac{LANET} can be employed to design intelligent transport systems, ensuring road safety \cite{v2lc,v2v2}. In \ac{V2V}, for example, a communication is established using head/tail lights as transmitters and photo-diodes/image sensors constituting receiver to provide reliable communication between vehicles. The urban infrastructures (traffic lights, street lights) can also be utilized for transmitting information related to the current circulation of traffic, vehicle safety, traffic information broadcast, and accident signaling. 

\begin{figure}[b]
\centering
\epsfig{file=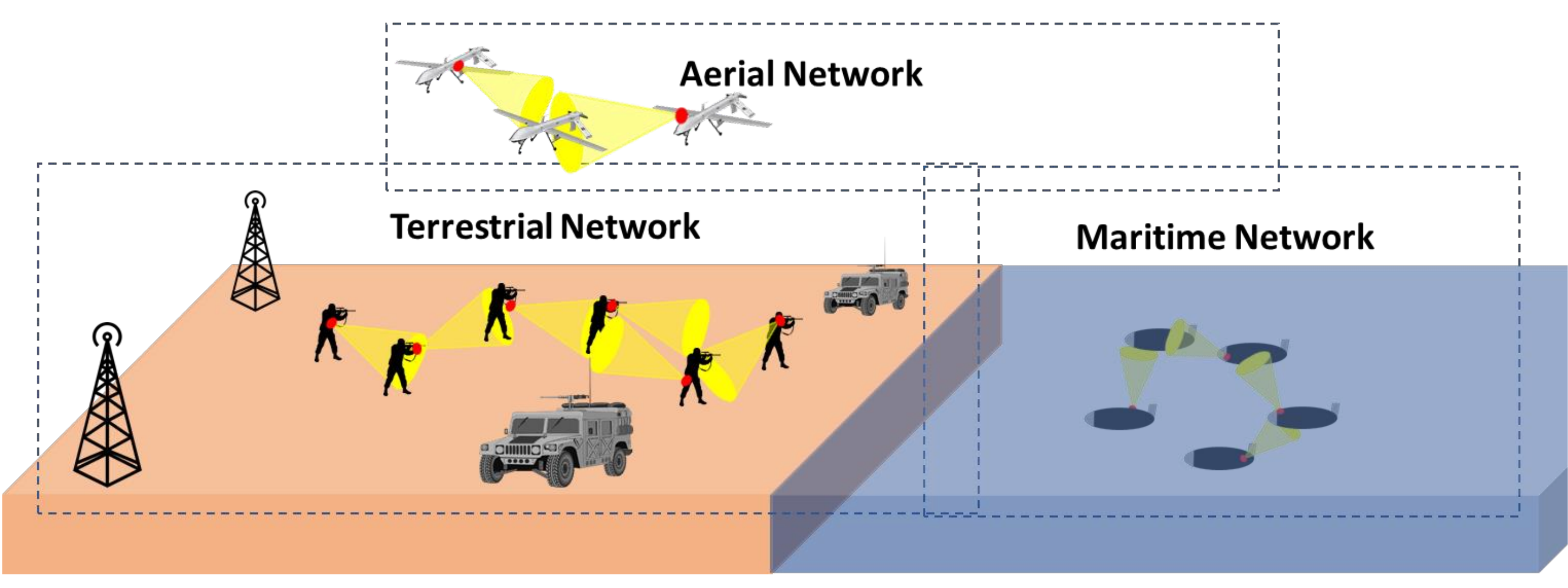, width=3.4 in,} 
\caption{Military application of LANET}\label{fig:Mil}
\end{figure}

\textbf{Military and Space Applications.} Tactical missions entailing the deployment of ships, soldiers, and unmanned surface vehicles across various operating environment including underwater, ground, and air can also leverage the use of \acp{LANET} as depicted in Fig. \ref{fig:Mil}. For example, we envision that self-organized autonomous underwater vehicles can form \ac{LANET} \cite{Chang_UW} to exchange high-data rate traffic via visible light carriers as a high-rate short-range alternative to acoustics. In ground applications, warfighters can self-organize in a \ac{LANET} in case of \ac{RF} interference and be connected to command; finally, in air/space LANETs, CubeSats or NanoSats can be connected to a satellite station via multihop \ac{VLC}. 

\begin{figure*}[t]
\centering
\epsfig{file=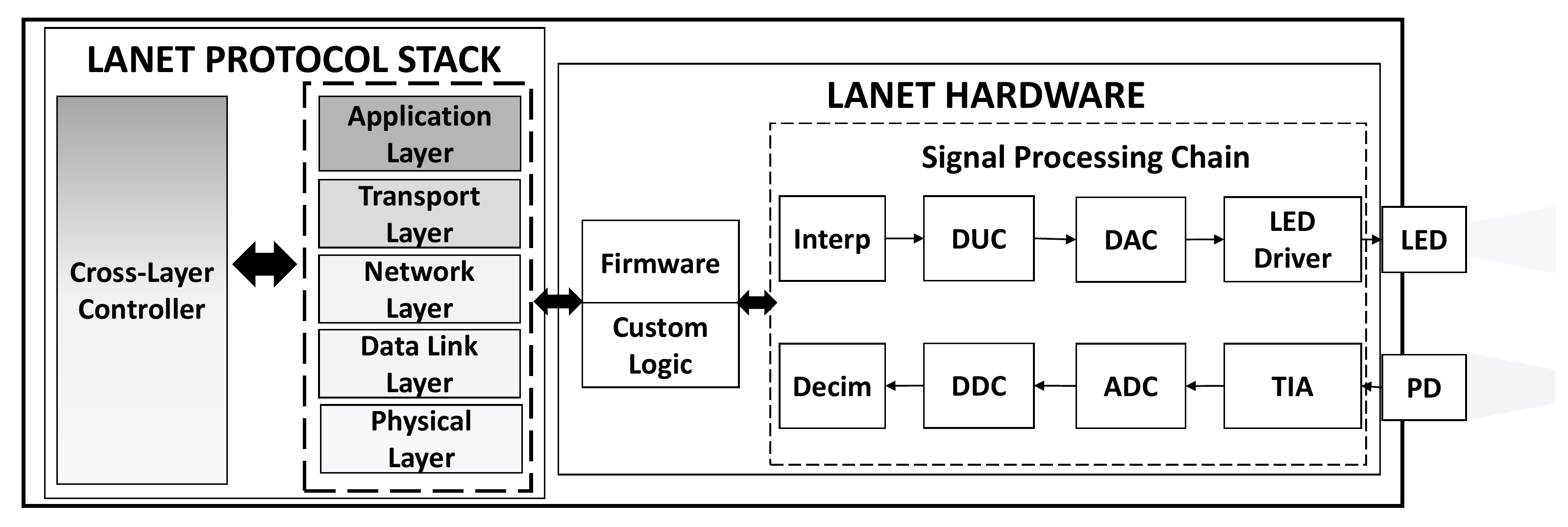, width=5.5 in,} 
\caption{Architecture of a cross-layer controller enabled LANET node}\label{fig:node}
\end{figure*}

Several of these applications will demand several hops beyond the tradition point-to-point links that are currently employed by \ac{VLC}. To make these a reality, significant work is required at the network layer to overcome some of the challenges specific to LANET like deafness (due to directionality) and blockage (due to the nature of propagation) that induces highly volatile route conditions. Therefore, it is evident that \textbf{reliability of route} and the opportunity to change the routing decision quickly will be the critical features distinguishing a routing protocol from traditional approaches. Previously, \textbf{cross-layer network optimization} has been explored in \ac{RF} networks \cite{Jagannath16GLOBECOM, Pompili06, Jagannath18TMC} but is especially crucial for \acp{LANET} to combat the volatile nature of links \cite{Bilgi14_xlayer, XuanLi_xlayer, AODV_V2LC}. Therefore, considering the above challenges, we can summarize the features essential for a routing protocol for LANETs as follows,

\begin{itemize}
\item A cross-layer framework as depicted in Fig. \ref{fig:node} is required to ensure collaboration of network layer with the data link layer to mitigate the degradation that is caused due to deafness and blockage and to maximize the probability of establishing full-duplex links.
\item Due to the highly dynamic nature of LANETs, an opportunistic routing protocol that uses a distributed algorithm to determine the optimal hops at each intermediate node in the multihop network is required.
\item Reliability of the routes each node can provide should be considered as a key metric while making routing decisions. The absolute channel condition itself may not be the best indicator of successful routes.
\end{itemize}

\section{Related Works}

The primary focus in \ac{VLC} has been to enable point-to-point communication with the goal of improving link data rates. Network layer protocols are usually derived from traditional methods to act as a facilitator of \ac{VLC} application \cite{OpenVLC} and hence \ac{LANET}-specific network layer design is still in its infancy. Nevertheless, we discuss some of the recent advancements at the network layer in this section. First, we begin by discussing how some of the challenges discussed earlier affect different types of routing techniques. 

\textbf{Proactive Routing}. Each node in the network maintains routing information for the entire network in a proactive (table-driven) routing protocol. This approach usually ensures lower end-to-end delays at the expense of larger overhead to maintain routes. Usually, in a traditional network with omnidirectional antennas, the nodes may use broadcast messages regularly to learn route changes. In a directional network, this becomes challenging and time intensive due to deafness and the need to exchange messages in every sector. This problem is further aggravated in \acp{LANET} due to the limited route lifetime due to varying link connectivity. Thus, there is a constant need to update routes but at the same time, it is extremely challenging and expensive to learn changes in the network. All these factors render it extremely difficult to maintain updated routing tables for the entire network.

\textbf{Reactive Routing}. In contrast to proactive routing protocols, a source node discovers route when it has to transmit a packet in a reactive routing protocol. This eliminates the need to maintain routing tables at every node and hence reducing the overhead and power consumption but may lead to higher delays. In the context of \acp{LANET}, it is difficult to discover all possible routes due to the narrow \ac{FOV} and without an adequate neighbor discovery scheme that overcomes blockage. Similarly, there is no guarantee that the route still exists once the route is discovered and the source starts transmitting packets towards the destination. Therefore, a route that theoretically provides the highest throughput but lacks alternate paths that might help route recovery in case of link failure may not be an ideal choice for \acp{LANET}. 
 
We have established how traditional design considerations may not be directly applicable to \acp{LANET}. Next, we look at some recent effort that has contributed towards the network layer of \acp{LANET}. In \cite{Shine}, authors propose a novel platform aimed at distributed multihop visible light communication that has $360$ degree coverage and is compatible with experimental boards such as Arduino, Beaglebone, Raspberry Pi. They identify the open problem in developing a multihop routing algorithm but do not propose a solution. Authors propose a greedy routing algorithm to support a fully wireless \ac{DCN} across racks in \cite{Qin_datacenter}. The greedy algorithm chooses the next hop such that it has the shortest distance to the destination. Their objective was to use \ac{VLC} to eliminate hierarchical switches and inter-rack cables, and thus reducing hardware investment, as well as maintenance cost. They were successful in showing an effective application of \ac{VLC} but the routing protocol does not consider any \ac{LANET} specific challenges.

There have been efforts to explore cooperative relaying \cite{Yang2013FullduplexRV, Yang2014FullduplexRV, Coop-relay} mainly for linear and triangular topologies in an indoor environment. In \cite{Yang2013FullduplexRV}, the authors propose a full-duplex relay system that uses a loop interference channel that is build on the channel model defined in \cite{IEEE_802_15_7}. The full-duplex relaying system is shown to outperform both half-duplex relaying and direct transmission especially while higher order modulation was employed. Another full-duplex relay for 3-luminaire triangle topology in indoor \ac{VLC} scenario is presented in \cite{Yang2014FullduplexRV} and \cite{Coop-relay}. In  \cite{Yang2014FullduplexRV}, the luminaries affixed on the ceiling next to each other tries to relay the source's transmission that reflects off the floor. The authors, proposed two approaches; (i) decode and forward and (ii) amplify and forward to accomplish cooperative relaying. Both these techniques effectively enable luminaries to extend its range of communication providing a wider coverage in indoor scenarios. Cooperative relaying can be used to enhance the link reliability and extend coverage but cannot be directly extended to enable routing in multihop networks. 

In \cite{AODV_V2LC}, the authors show that improved end-to-end delivery ratio can be achieved by using multihop broadcast that accounts for the intermittent blockage problem of \ac{VLC} links in vehicular visible light communication (V2LC) networks. The need for a routing protocol specifically designed for \acp{LANET} has been identified in their work but was considered out of the scope of their objective. 

A hexagonal cylindrical design to provide omnidirectional access to directional \ac{VLC} network is proposed in \cite{ZEYUWU12}. Each face in the hexagonal design has IR transmitters and \ac{PD} receivers to provide omnidirectional access. The author designs methodology to avoid the sudden blockage by finding alternate paths to the intended destination. Accordingly, when the base station (intended for the ceiling) or the user device (intended for the desk) loses connection it first checks if a connection can be established using any other faces. If that fails, source checks if a previously known route exists and sends validate packet if it does. If such route does not exist, the source sends reactive route discover packet with preset forward depth count looking for a node which has the path to the destination node. If in a given period of time (associated with forward depth count) there is no response from any node, they consider that there is no such node. While the proposed solution aims to mitigate the effects of blockage, the constant disconnect and route discovery may cause excessive delays deteriorating the overall network throughput. 
 
  
In the light of these recent efforts, it is clear that the feasibility of \ac{LANET} depends on a dynamic routing protocol that establishes reliable routes from source to destination. In order to ensure this reliability in the presence of blockage, deafness, and to utilize the inherent full-duplex nature of \ac{VLC}, a cross-layer approach between the data link and the network layer is essential.   

To the best of our knowledge, this is the first work that addresses the challenges in enabling distributed and dynamic routing algorithm specifically for \acp{LANET}. Accordingly, the contributions of this work are as follows,
\begin{itemize}
\item We propose a cross-layer routing algorithm that interacts closely with the \ac{MAC} layer to ensure opportunistic packet forwarding thereby mitigating effects of deafness, blockage and hidden node problem.
\item Due to the volatile nature of routes in \ac{LANET}, we establish the requirement of highly dynamic routing technique and hence design a routing algorithm where hop-by-hop decision making is employed to ensure adaptability.
\item We formulate a reliability score that enables complete decentralized operation of the network with the objective to maximize the expected throughput of the network. 
\item  Extensive simulations are conducted to analyze the behavior of the proposed routing algorithm in various dynamic operating environment.
\end{itemize}

\section{System Model}\label{sec:system_model}

\begin{figure*}[!t]
\minipage{0.3\textwidth}\hspace{.8 cm}
\includegraphics[width=1.3 in]{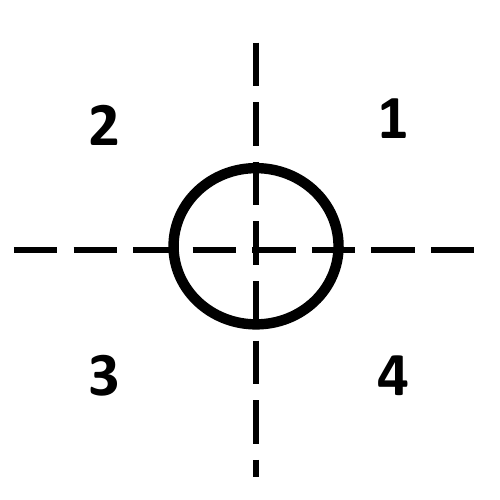}\vspace{-0.24cm}
\caption{Sectors}\label{fig:sectors} 
\endminipage\hfill
\minipage{0.69\textwidth}
\includegraphics[width=5 in]{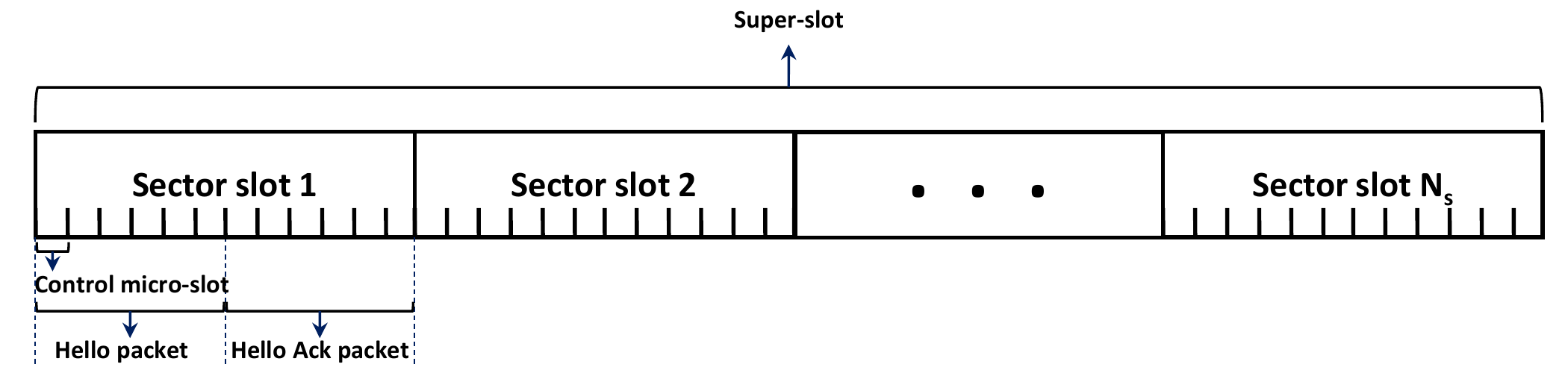}\vspace{.4cm}
\caption{Super-slot structure}\label{fig:SuperSlot} 
\endminipage\hfill
\end{figure*}

We envision that operating scenarios for \ac{LANET} will involve a dense sensor network that carries a large amount of information from the sensors to sinks using low range, high data rate links. For example, these sinks can represent data center in \acp{VANET}, command and control centers in military applications etc. In this paper, we assume multiple low cost \acp{VLN} are densely distributed with the objective to collect sensing information and transfer it to one of the sinks deployed throughout the network. To this end, consider a multihop \ac{LANET} with  $N$ static \acp{VLN} modeled as a directed connectivity graph $\mathcal{G}(\mathcal{U},\mathcal{E})$, where $\mathcal{U}=\{u_0,u_1, ... , u_{j}\}$ is a finite set of \acp{VLN} and $\mathcal{U}=\{u_{j+1}, ... , u_{N}\}$ is the set of sinks of the graph, and $(i,j)\in \mathcal{E}$ represents a feasible unidirectional wireless link from node $u_i$ to node $u_j$ (for simplicity, we also refer to them as node $i$ and node $j$) representing neighboring relationships, i.e., a feasible link exists if the nodes are close enough. In LANET, each node consists of \ac{LED} luminaries and \acp{PD} adopted as transmitters and receivers, respectively. Since the transmissions are directional, the directions to which the \ac{FOV} of each node can be set to are represented by $N_s$ equal sectors $s \in \mathcal{S}$ as shown in Fig. \ref{fig:sectors}. The FOVs of typical \acp{LED} and \acp{PD} can vary from $\pm 10^{\circ}$ to $\pm 60^{\circ}$ \cite{web1, PrinciplesLED}, e.g. Vishay TSHG8200, OSRAM LCW W5SM Golden Dragon and Vishay PD TESP5700. Here, for the sake of simplicity, but without loss of generality, we choose \ac{FOV} for both \ac{LED} and \ac{PD} to be $\pm 45^{\circ}$, leading to four sectors. This can be easily extended accordingly to the FOV of the hardware available on specific \ac{VLN}. We also assume that a \ac{VLN} is capable of directing its \ac{FOV} to all the $N_s$ sectors when required for transmission and listening. This is possible with multiple \acp{LED} and \acp{PD} \cite{ZEYUWU12} that can be used depending on the sector to which nodes require access. The neighbors are grouped into sectors based on their location. Thus, the superset of neighbors for node $i$ consists of the set of neighbors in each sector represented as $\mathcal{NB}^i \in \{\mathcal{NB}^i_1, \mathcal{NB}^i_2,..., \mathcal{NB}^i_{N_s}\}$, where $\mathcal{NB}^i_{s}$ denotes the neighbors of node $i$ in sector $s$ given by $\mathcal{NB}^i_{s} \triangleq \{j: (i,j) \in \mathcal{E}\} $. 

Let the traffic in the network consist of multiple sessions $q=1,2,...,Q$, characterized by the source-sink pairs. The goal of this work is to optimize the expected throughput of the network while taking into consideration the reliability of the routes due to specific challenges associated with \acp{LANET}. In this work, a feasible next hop for a session is defined as any neighbor that is closer to the destination and is termed as forward progress. In this context, each session $q$ in node $i$ belongs to one or more sector queue sets $q \in \mathcal{Q}_i^s$ such that the sector contains neighbors that ensure forward progress for packets in a queue. This information will be used by the \ac{VLN} while choosing an optimal sector to forward packets. It is important to understand that a session $q$ can be a component of more than one sector queue sets because the packet of a session can achieve forward progress through more than one sector. 
The \acp{VLN} in the network are assumed to be synchronized with each other using techniques like Global Positioning System (GPS) based clock synchronization. The time spent listening to each sector is called sector duration ($t_{sec}$) and this forms a sector slot as shown in Fig. \ref{fig:SuperSlot}. The sector slot is further divided into multiple control micro-slots (CMS). Control packets are transmitted only at the beginning of a CMS. The duration of a CMS is set such that the transmission of a control packet can be completed in one CMS. A set of $N_s$ sector slots form a super-slot. \acp{VLN} have two operational states; \ac{S-IDLE} and \ac{TR}. In \ac{S-IDLE}, nodes sequentially listen in each sector following a fixed pattern. In this way, a \ac{VLN} that has to transmit in a given sector knows the appropriate sector slot when the idle neighbors (in the given sector) will be listening, thus mitigating the effect of deafness. The channels used by the LANET are divided into independent \ac{CC} and \ac{DC}.

\section{Formulation of Routing Algorithm}

Most routing algorithms in \ac{RF} based ad hoc network are designed to optimize network parameters like throughput, delay, energy consumption etc. These cannot be the sole metric of consideration in \ac{LANET} due to its unique challenges discussed earlier. The link state in \acp{LANET} are highly dynamic and can be interrupted due to blockage or deafness in addition to channel conditions itself. \emph{Therefore, route reliability becomes a key metric for consideration while designing routing algorithm for LANET.} The reliability of a route can be defined as the probability of successfully delivering a packet from $i$ to the desired sink $k$ on the first attempt. It is given as follows,
\begin{equation}
p_{r}(i:k)= \prod_{(i,j) \in \mathcal{L}_r }p(i,j) \label{eq:p_r}
\end{equation}

where $\mathcal{L}_r$ is the set of all links $(i,j)$ in route $r$, $p(i,j)$ is the probability that packet is successfully forwarded from $i$ to $j$ in the first attempt. The value of $p(i,j)$ depends on packet error probability ($p_e$),  probability of blockage ($p_{ij}^b$), and probability of $i$ winning the contention to establish link with $j$ in the first attempt ($p^{acs}_{ij}$). Therefore, $p(i,j)$ can be represented as follows,
\begin{equation}
p(i,j)=(1-p_e).p^{acs}_{ij}.(1-p_{ij}^b)
\end{equation}

where $p^{acs}_{ij}$ denotes the probability that node $i$ can negotiate access to node $j$ which in turn depends on number of nodes ($M$) in set $\mathcal{NB}^j_{s}$ where $s$ is the sector to which $i$ belongs. Assuming worst case scenario where every node has a packet to transmit and each node chooses a random backoff value between the range 
$(0, CW-1)$ where $CW$ is the contention window size, it can be represented as,

\begin{equation}
p^{acs}_{ij}=p_0 (1-p_0)^{M-1}
\end{equation}

such that $p_0$ is the probability that a \ac{VLN} transmits in a timeslot and is given by \cite{Bianchi_CSMA},
\begin{equation}
p_0=\frac{2}{CW+1}
\end{equation}

Using (\ref{eq:p_r}), the probability of delivering packets from $i$ to $k$ over at least one of the possible routes can be given by,
\begin{equation}
p(i:k)= 1 - \prod_{r \in \mathcal{R}^i_{k} }\left[1-p_{r}(i:k)\right] \label{eq:p(i:k)}
\end{equation}
where $\mathcal{R}_k^i$ is the set of all possible routes from node $i$ to sink $k$ (we only consider routes in which each hop makes some forward progress towards $k$). Therefore, the expected throughput of a session $q$ can be defined as follows,
\begin{equation}
E[T(q)]=p(i:k).T(q)
\end{equation}
where $T(q)$ is the maximum achievable throughput for a session $q$ from node $i$ to $k$. The overall objective of the proposed routing algorithm for LANET is defined as shown below,

\begin{align}
				\textup{Maximize}\!&:  \sum_{q \in \mathcal{Q}} E[T(q)]\\
				\textup{subject\; to}\!&:\notag \\
				& \text{Link capacity constraint} \\
                & \text{Maximum queue size constraint}\\
                & \text{Power budget constraint}
\end{align}

Computing $p(i:k)$ requires global knowledge of the network in order to consider all possible routes and link probabilities from node $i$ to $k$. Therefore, in this paper, to enable distributed operation, we define a \ac{RRS} based on approximation of $p(i:k)$ for each node $i$ to indicate a measure of expected success in reaching the sink $k$ through $i$. Each node will use the \ac{RRS} of its immediate neighbors to determine its routing strategy at each hop. Accordingly, we define \ac{RRS} of node $i$ with respect to sink $k$ as follows,
\begin{equation}
\Gamma_i^k= \beta_i^k . \left[1 - \min_{s \in \mathcal{S}} \left(\prod_{j \in \mathcal{NB}_s^i/h_i^k<h_j^k }\left[1-p(i,j).\Gamma_j^k\right]\right)\right] \label{eq:score}
\end{equation}
where,  
\begin{equation}
\beta^k_i=\frac{b_{max}-b^i_k}{b_{max}}
\end{equation}

where $b^i_k$ is the total number of packets destined for $k$ backlogged at $i$ and $b_{max}$ represents the maximum buffer size. $\beta^k_i$ is meant to penalize the score of $i$ w.r.t particular sink $k$ when the node is heavily backlogged with packets intended for $k$. The absence of such penalization may lead all traffic to follow the same route regardless of congestion. \emph{The use of $\beta^k_i$ aims to ensure a balance between reliability and congestion.} Next, $h_i^k$ denotes the \ac{MHC} from $i$ to $k$. It can be seen in the formulation of (\ref{eq:score}) that only neighbors with lower \ac{MHC} to the respective sink $k$ contributed towards the \ac{RRS}. Since \ac{VLN} can operate on only one sector at a time, this should be reflected in the calculation of \ac{RRS}. Therefore, the final value of \ac{RRS} of any given node corresponds to the sector that provides the highest value among all sectors in $\mathcal{S}$. Looking closely at the formulation of \ac{RRS}, one realizes that this distributed estimation still reflects the overall structure of (\ref{eq:p(i:k)}) along with the addition of $\beta^k_i$.  More importantly, \emph{it can be seen from the above definition of \ac{RRS} that every node can calculate its score with respect to each sink using information gathered from immediate neighbors.} This critically ensures a scalable and distributed operation.

Initially, each \ac{VLN} sets its \ac{RRS} to zero and \ac{MHC} to infinity for each sink in the network. Thereafter, \acp{VLN} listen to neighbor's control packets to compute \ac{RRS} and \ac{MHC}. The first set of \ac{RRS} is calculated by \acp{VLN} within one hop from the sinks from the overheard control packets transmitted by the sink. In this case, for a node $x$ one hop away from sink $k$, \ac{RRS} is simply $\Gamma_x^k=p(x,k)$ and updates its \ac{MHC} to sink $k$ as $h_x^k=1$. Subsequently, the computed \ac{RRS} is appended to the control packets that it transmits. In this manner, whenever a node $i$ receives an updated $\Gamma_j^k$ from neighbor $j$ (that provides forward progress w.r.t $k$), $i$ updates $\Gamma_i^k$ (according to eq. \ref{eq:score}) and $h_i^k$. In this manner, each node keeps updating its \ac{RRS} and \ac{MHC} w.r.t each known sink in the network. 

As discussed earlier, we realize that a cross-layer approach is required to combat unique challenges posed by LANETs. The data-link layer and network layer need to coordinate with each other to optimize the network performance. Therefore, we design a cross-layered routing algorithm that embeds the \ac{RRS} into an opportunistic \ac{MAC} protocol designed specifically for \acp{LANET}. To this end, we significantly extend the \ac{MAC} protocol, VL-MAC \cite{Jagannath18ICNC} to interact with the network layer and include optimized routing decisions while negotiating the access of the medium. The concept of opportunistic link establishment was first introduced for \acp{LANET} in \cite{Jagannath18ICNC}.  




\begin{figure*}[h]
\centering
\includegraphics[width=5.5 in]{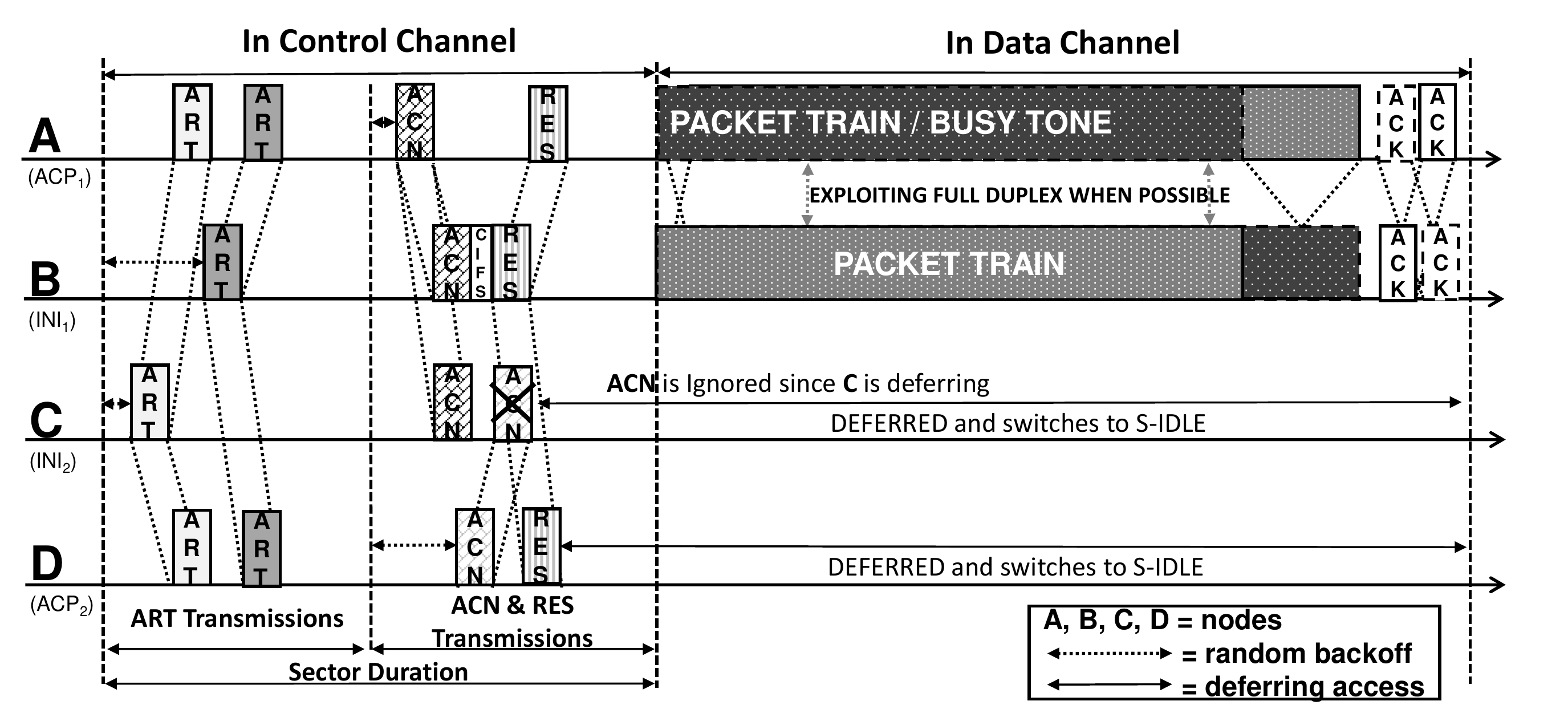}\vspace{ 0 cm}
\caption{Timing diagram}\label{fig:timdia} 
\end{figure*}

The timing diagram of the mechanism followed by VL-MAC is depicted in Fig. \ref{fig:timdia}. Since \ac{VLC} inherently supports full-duplex communication, to avoid confusion, we replace the terms transmitter and receiver by initiator and acceptor depending on which node initiates communication. Consider four nodes $A$, $B$, $C$ and $D$, among which $B$ and $C$ are the initiators with packets to be transmitted and $A$ and $D$ are prospective acceptors in \ac{S-IDLE}.  Once a node has packets to transmit, it has to choose a sector to transmit such that it maximizes the initiator's utility function ($U^i_{ini}(s)$). Consider a node $i$ with packets intended for sink $k$, and let $j$ be the possible next hop. The initiator's utility function for node $i$ is given by,

\begin{equation}
U^i_{ini}(s)=\sum_{q \in \mathcal{Q}^i_s} \sum_{j \in \mathcal{NB}^i_s}  b^i(q) \;\overline{d_{ij}^k} \;\overline{\Gamma_j^k} \label{eq:INI}, \;\; \forall j:d_{ik}- d_{jk}>0\\
\end{equation}
where, 
\begin{equation}
\overline{\Gamma_j^k}=\frac{\Gamma_j^k}{ \max_{j \in \mathcal{NB}^i}[\Gamma_j^k]} \text{ and } \overline{d_{ij}^k}=\frac{d_{ik}- d_{jk}}{d_{ij}} \label{eq:dbar},
\end{equation}
where $b^i(q)$ is the backlog length of session $q \in \mathcal{Q}_i^s$ at node $i$, $\mathcal{Q}^i_s$ is the set of all sessions with packets that can be forwarded through sector $s$. The measure of forward progress is provided by $\overline{d_{ij}}$ \cite{Jagannath18ICNC}. The normalized \ac{RRS}, $\overline{\Gamma_j^k}$ provides higher utility to the sector with neighbors that provide more reliable routes. It is critical to understand why $\overline{\Gamma_j^k}$ is used over just $\Gamma_j^k$. Investigating (\ref{eq:score}) closely, it can be seen that $\Gamma_j^k$ generally decreases as the number of hops to sink increases. Therefore, using $\Gamma_j^k$ instead of $\overline{\Gamma_j^k}$ would give unfair advantages to a session whose destination is closer to $i$. The goal of $\overline{\Gamma_j^k}$ in (\ref{eq:INI}) is to ensure that neighbor that provides relatively higher RRS w.r.t a given sink contributes to a larger utility value. The summation over all feasible neighbors ensures that the utility function increases proportionally to the number of feasible neighbors in the given sector, which in turn increases the probability of finding an available next hop mitigating the effect of deafness.    

The goal of this utility function is to introduce the concept of opportunistic link establishment in contrast to traditional methods where a forwarding node is chosen before the negotiation for channel access begins. This mitigates the inaccessibility caused due to blockage or deafness. Accordingly, $i$ chooses the optimal sector $s^{*}_i$ that maximizes its utility function and can be represented as,
\begin{equation}
s_i^*=arg\;max_{s \in \mathcal{S}}\left(U^i_{ini}(s)\right)
\end{equation}

Accordingly, in this example, lets assume $B$ and $C$ choose the same sector which corresponds to their maximum $U_{ini}(s)$. Nodes $B$ and $C$ choose a random backoff depending on their $U_{ini}$ and broadcast an \textit{\ac{ART}} packet if the channel is idle within the \textit{\ac{ART}} transmission period of the sector duration. The \textit{\ac{ART}} consists of the information regarding the source node (initiator) such as node ID, location, \ac{RRS}, backlog length of all sessions considered for the given sector and channel state. As shown in Fig. \ref{fig:timdia}, both $A$ and $D$ listen to control packet during the corresponding sector duration. On reception of \textit{\acp{ART}}, $A$ and $D$ will switch to \ac{TR} and calculate their respective acceptor's utility function, $U^j_{acp}(i)$, using information from all the \textit{\acp{ART}} received during the sector duration. The $U^j_{acp}(i)$ for any initiator-acceptor pair $i$ and $j$ can be computed as follows,
\begin{equation}
U^j_{acp}(i)= \left[\eta_{ij}(q^*_i) C_{ij} \right] +\left[\eta_{ji}(q^*_j)C_{ji}\right] \label{eq:ACP}
\end{equation}
where $q^*_i$ is the session selected for transmission from $i$ to $j$ such that it maximizes the weighted differential backlog given as follows, 
\begin{align}
\eta_{ij}(q^*_i)=arg\;max_{q \in \mathcal{Q}^i_s} \left[\overline{\Gamma_j^k}\;\overline{d_{ij}^k}\;\left(b^i(q) - b^j(q)\right) \right]
\end{align}
It can be seen that (\ref{eq:ACP}) includes the product of maximum weighted differential backlog and channel capacity ($C_{ij}$) in both directions as defined in \cite{Jagannath18ICNC}. This implies the initiator-acceptor pair that can achieve  higher combined throughput using full-duplex communication gets access to the channel thereby improving the overall throughput of the network. It is important to note that in contrast to \cite{Jagannath18ICNC}, the utility functions include the \ac{RRS} that governs the routing decision and hence will implicitly lead to reliable routes and higher throughput. 

According to the above discussion, $A$ and $D$ choose the initiator ($B$ or $C$) that they want to provide access. The acceptors also select the initiator's session and acceptor's session for full-duplex communication such that it maximizes their respective $U^j_{acp}(i)$ as shown below,
\begin{equation}
(i^{*},q^*_i,q^*_j)= \text{arg max}\left(U^j_{acp}(i)\right).
\end{equation}
This is the second critical step taken by the cross-layer routing protocol \textit{to maximize the expected network throughput by choosing initiator-acceptor pairs favoring opportunities for establishing full-duplex communication.} These chosen parameters are encapsulated in a \textit{\ac{ACN}} packet and transmitted by the acceptors to the chosen initiators. In  this particular example, after a $U_{acp}$ dependent random backoff, $A$ transmits an \textit{\ac{ACN}} to $B$. The \textit{\ac{ACN}} contains information that is used by the initiator to set the transmission parameters (modulation, power, and channel if applicable). Accordingly, $B$ receives the \textit{\ac{ACN}} from $A$  and $C$ overhears this \textit{\ac{ACN}} intended for $B$. Next, $B$ transmits \textit{\ac{RES}} packet to reserve the time required to complete the transmission. 
Node $C$ learns that it was not chosen for transmission by overhearing the \textit{\ac{ACN}}, and hence defers access and returns to the \ac{S-IDLE}. Similarly, $D$ overhears the \textit{RES} packet and returns to \ac{S-IDLE}.

After this three-way handshake, nodes $A$ and $B$ perform full-duplex data transmission as depicted in Fig. \ref{fig:timdia}. The respective receivers transmit the \textit{\ac{ACK}} packet after the reception of the data packet. After the completion of the full-duplex transmission, both the nodes return to the \ac{S-IDLE}.  In cases where there is no opportunity for full-duplex communication (acceptor does not have any session to be transmitted to the initiator), a busy tone is transmitted by the acceptor. This is to ensure that other nodes sense the channel to be busy from both directions of the initiator-acceptor pair and reduce the problems associated with hidden node. All these factors collectively mitigate the effects of deafness, blockage and hidden node problem while favoring the establishment of full-duplex links. These factors along with the carefully designed \ac{RRS} based route selection strategy maximizes the expected throughput of the network.

\section{Simulation}

\begin{figure*}[h!]
\minipage{0.5\textwidth}\hspace{-0.1 cm}
\centering
\includegraphics[width=3.3 in]{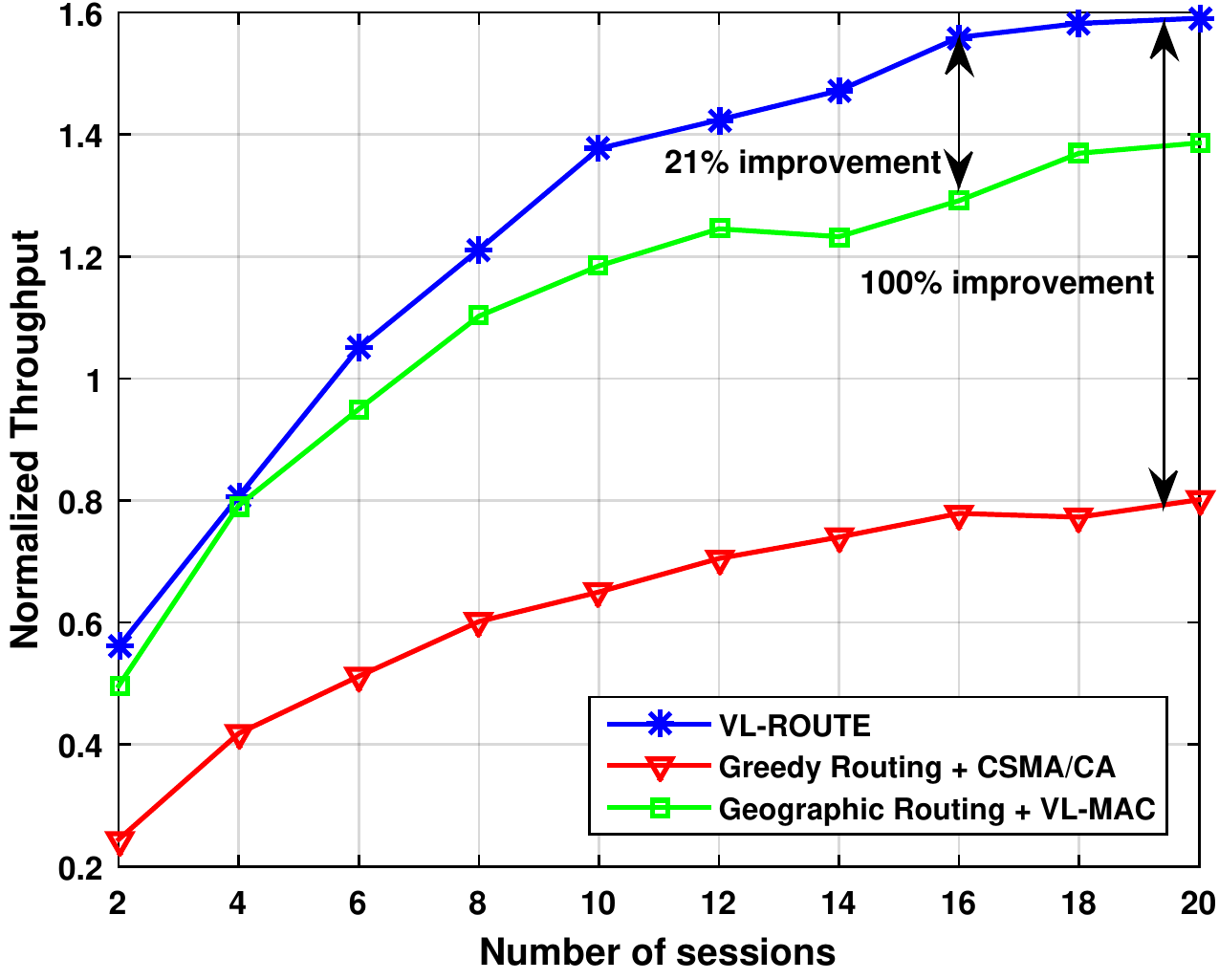}\vspace{ 0 cm}
\caption{Throughput vs No. of sessions}\label{fig:Th_grid} 
\endminipage\hfill
\minipage{0.5\textwidth}\hspace{-0.7 cm}
\centering
\includegraphics[width=3.3 in]{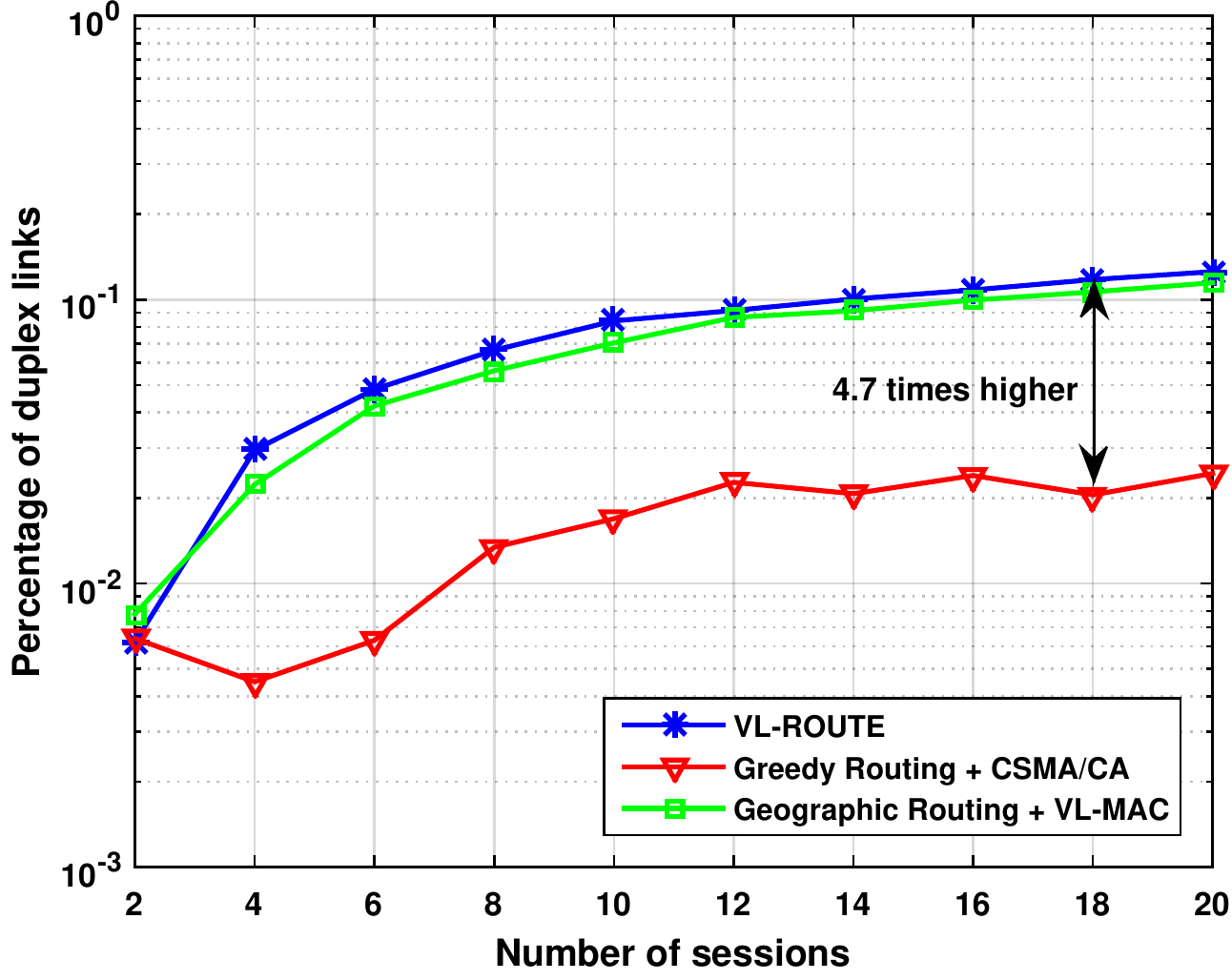}\vspace{ 0 cm}
\caption{Full-duplex link vs No. of sessions}\label{fig:DP_grid} 
\endminipage
\end{figure*}

To evaluate the performance of VL-ROUTE, we implement a packet level simulator that operates at the network layer but interacts closely with the data link layer. The simulator only considers packet loss caused due to collisions and channel condition (i.e. based on the packet error rate of each link). This framework can be easily extended to include any modulation and coding scheme at the physical layer and will show a similar trend in performance at the network layer. The simulator is used to compare the performance of the proposed VL-ROUTE with a greedy routing that employs \ac{CSMA/CA} based \ac{MAC} (for simplicity we refer to this as GR-CSMA). The greedy routing is similar to \cite{Qin_datacenter}, a \ac{VLN} that has a packet to transmit choose a neighbor that is closest to the intended sink to forward the packet. To ensure a fair comparison, \acp{VLN} are synchronized in both cases and perform full-duplex communication links whenever possible. The network consists of $100$ \acp{VLN} with a transmission range of $4\;\mathrm{m}$ and each session in the network is characterized by the source node and an indented sink. The size of control and data packets were $20\;Bytes$ and $2500\;Bytes$ respectively and the data rate was set to $10\;Mbps$. 

\begin{table}[h]
\small
  \caption{Parameters of simulation in Grid Topology}
  \label{tb:Grid}
  \centering
  \begin{tabular}{ll}
    \toprule
    \textbf{Parameters}    & \textbf{Values} 		\\
    \midrule
    Size                     &  $25\;m$ $\times$ $25\;m$			\\
    Mean packet error rate   &  $0.2$          \\
    Number of sessions       &  $2$ to $20$          \\
    Total Packets per session&  $200$          \\
    Number of Sinks          &  $5$            \\
    \bottomrule
  \end{tabular}
\end{table}
\textbf{Grid Topology.} VL-ROUTE is first evaluated in a fully connected $10$ $\times$ $10$ grid network using the parameters shown in Table \ref{tb:Grid}. To perform a rigorous evaluation, in addition to GR-CSMA, we compare the VL-ROUTE to VL-MAC. Though VL-MAC was designed in order to facilitate cross-layered operation, it is important to recognize that VL-MAC by itself cannot serve as stand-alone routing algorithm. This is because VL-MAC by itself does not have the complete mechanism to determine existing routes to the sink. Whereas, in VL-ROUTE, the presence of a neighbor $j$ with a non zero $\Gamma_j^k$ indicates the presence of route to the sink $k$. But here, to perform a thorough evaluation of the proposed VL-ROUTE, we devise a way to compare it to VL-MAC. To this end, we assume all the \acp{VLN} using VL-MAC know the location of the sinks and hence uses VL-MAC in each hop to perform a geographical routing to the sink and ensures forward progress in each hop. 

As discussed earlier, blockage is one of the most critical challenges of LANET. To simulate this behavior, we randomly set $25\%$ of the links to have a $90\%$ chance of blockage and the remaining $75\%$ of the links to have $5\%$ chance of blockage. In real life scenarios, this would be the difference between a busy walkway or corridor that has a high probability of blockage versus most parts of the building that might receive considerably less foot traffic or obstruction. Later, we will evaluate how varying blockage levels impact VL-ROUTE in more detail. Both $p_e$ and $p_{ij}^b$ can be estimated by monitoring the previous activity on the given link ($p_e$ will depend significantly on modulations and coding used by the physical layer) with some estimation error associated with it. In this first simulation, we set the estimation error to $5\%$ but provide further analysis of the effects of estimation error later.

\begin{figure*}[t]
\minipage{0.5\textwidth}\hspace{-0.7 cm}
\centering
\includegraphics[width=3.3 in]{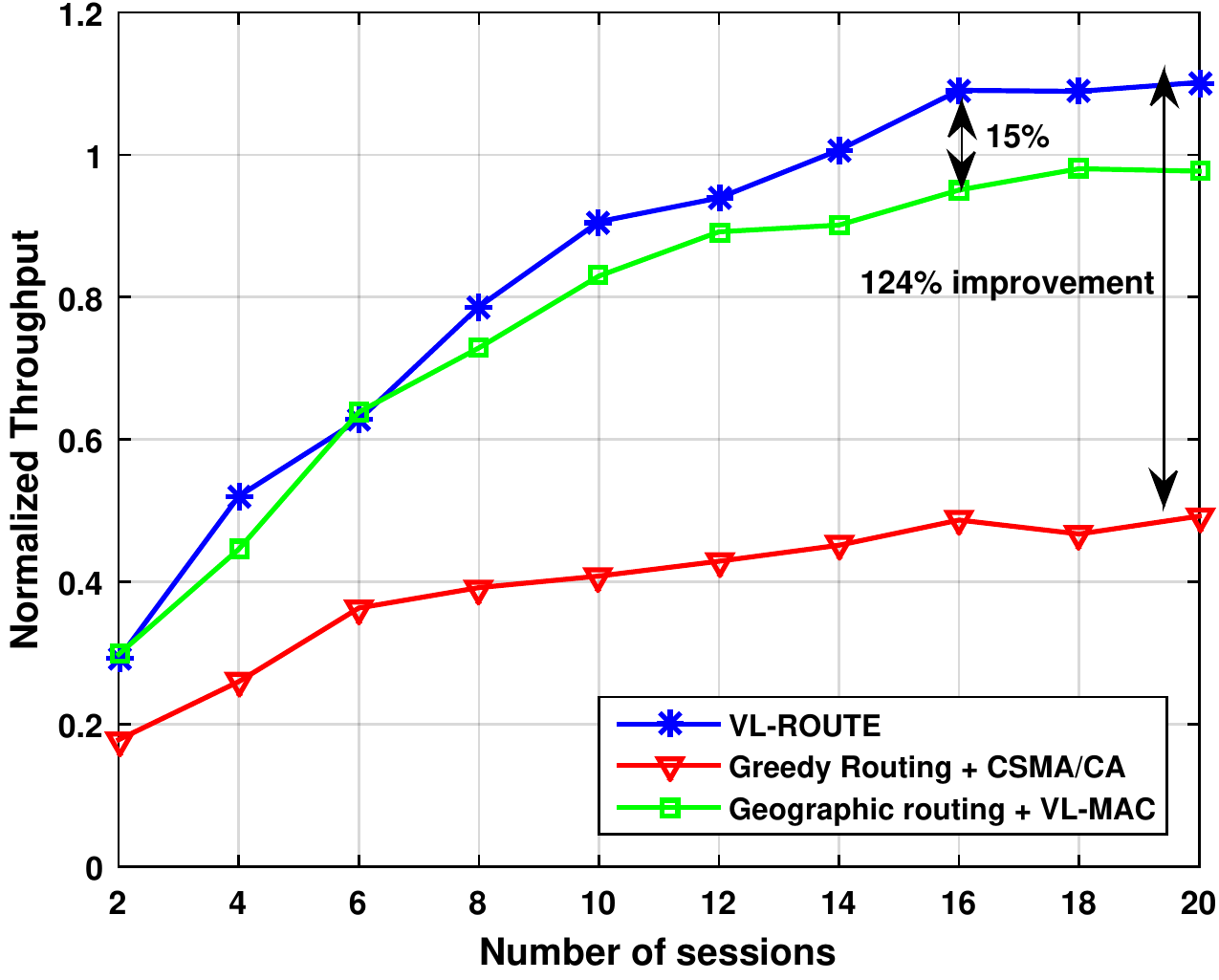}\vspace{ 0 cm}
\caption{Throughput vs No. of sessions}\label{fig:Th_rand} 
\endminipage
\minipage{0.5\textwidth}\hspace{-0.1 cm}
\centering
\includegraphics[width=3.3 in]{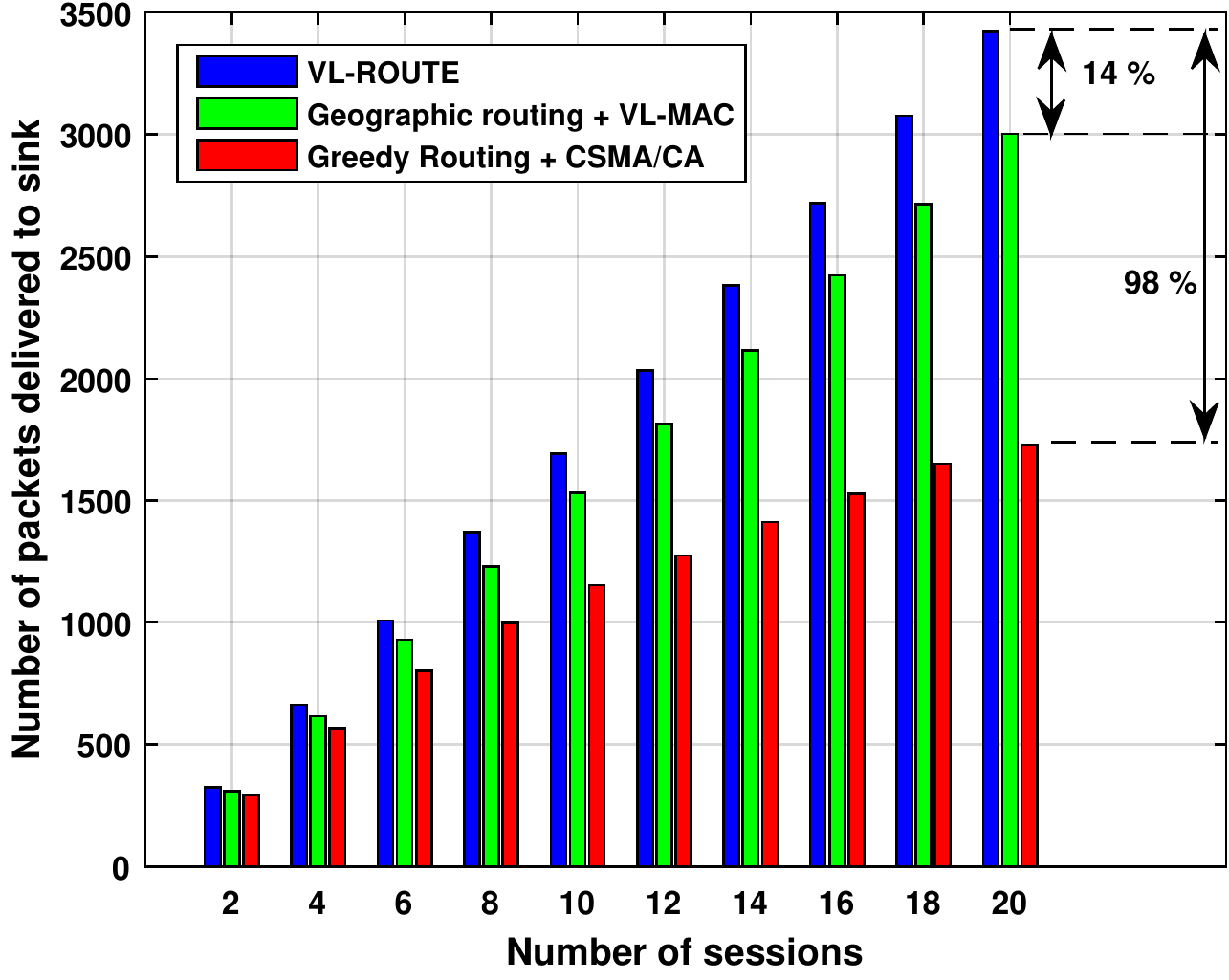}\vspace{ 0 cm}
\caption{Throughput vs No. of sessions}\label{fig:Pack_rand} 
\endminipage\hfill
\end{figure*}



First, we compute the throughput normalized to link rate as the number of sessions in the network increases. As seen in Fig. \ref{fig:Th_grid}, VL-ROUTE takes into account the unique characteristics of \ac{LANET} and outperforms traditional approach employed by GR-CSMA by up to $100\%$. This improvement in network throughput can be attributed to three main reasons; (i) consideration of route reliability in a distributed manner encourage packets to select route that provides the least resistance (caused by blockage or unfavorable channel conditions) to intended sink (ii) the cross-layer interaction with the link layer that provides opportunistic link establishment mitigates the effects due to deafness, blockage, and hidden node, and (iii) combining \ac{RRS} with the remaining factors of VL-MAC maximizes the probability of establishing full-duplex links while choosing optimal routes. The influence of the novel design of VL-ROUTE can be further substantiated by comparing the performance of VL-ROUTE to the network that used geographic routing (which assumes the location of the sink is known) with VL-MAC. This improvement in performance can be primarily attributed to route choices since the ratio of full-duplex (see Fig. \ref{fig:DP_grid}) links are similar for both VL-MAC and VL-ROUTE (although both are much higher with compared to GR-CSMA).

\textbf{Random Topology.} Since the grid network is uniform deployment and has a uniform neighborhood, we simulated a random topology to evaluate the performance of VL-ROUTE in a non-uniform deployment. Therefore, Fig. \ref{fig:Th_rand} shows that even in a random deployment the proposed algorithm outperforms GR-CSMA and VL-MAC because not only is it able to identify all the feasible hops but can also direct traffic in such a way that it maximizes the expected throughput of the network. VL-ROUTE achieves up to $124\%$ improvement in throughput with respect to GR-CSMA and achieves  $15\%$ improvement w.r.t VL-MAC. Figure \ref{fig:Pack_rand} depicts the number of packets that were successfully delivered to the sink by each routing algorithm. Since the random topology does not guarantee a fully connected network there could be several dead-end paths. The construction of \ac{RRS} score ensures the maximum delivery of packets while the opportunistic \ac{MAC} protocol itself perform reasonably well. Overall, VL-ROUTE delivers up to $98\%$ and $14\%$ more packets than GR-CSMA and VL-MAC respectively.



\textbf{Blockage Analysis.} Next, to study how VL-ROUTE is affected due to various levels of blockage that might be encountered in real-life implementation, we simulate various levels of blockage scenarios. To this end, we use the same parameters as in the random topology and vary the percentage of nodes experiencing $90\%$ blockage from $0$ to $60\%$. All other nodes in each test point will be set to experience $5\%$ chance of blockage. The number of sessions in the network is set to $5$. As expected, the throughput decreases as the percentage of nodes experiencing severe blockage increases but in all the scenarios VL-ROUTE outperforms GR-CSMA. The improvement is minimum ($69\%$) at the lowest level of blockage and increases up to $114\%$ for higher levels of blockage. This proves how VL-ROUTE can adapt to any level of blockage and provide optimal performance in any given scenario while operating in a distributed manner.

\begin{figure}[h]
\centering
\includegraphics[width=3.3 in]{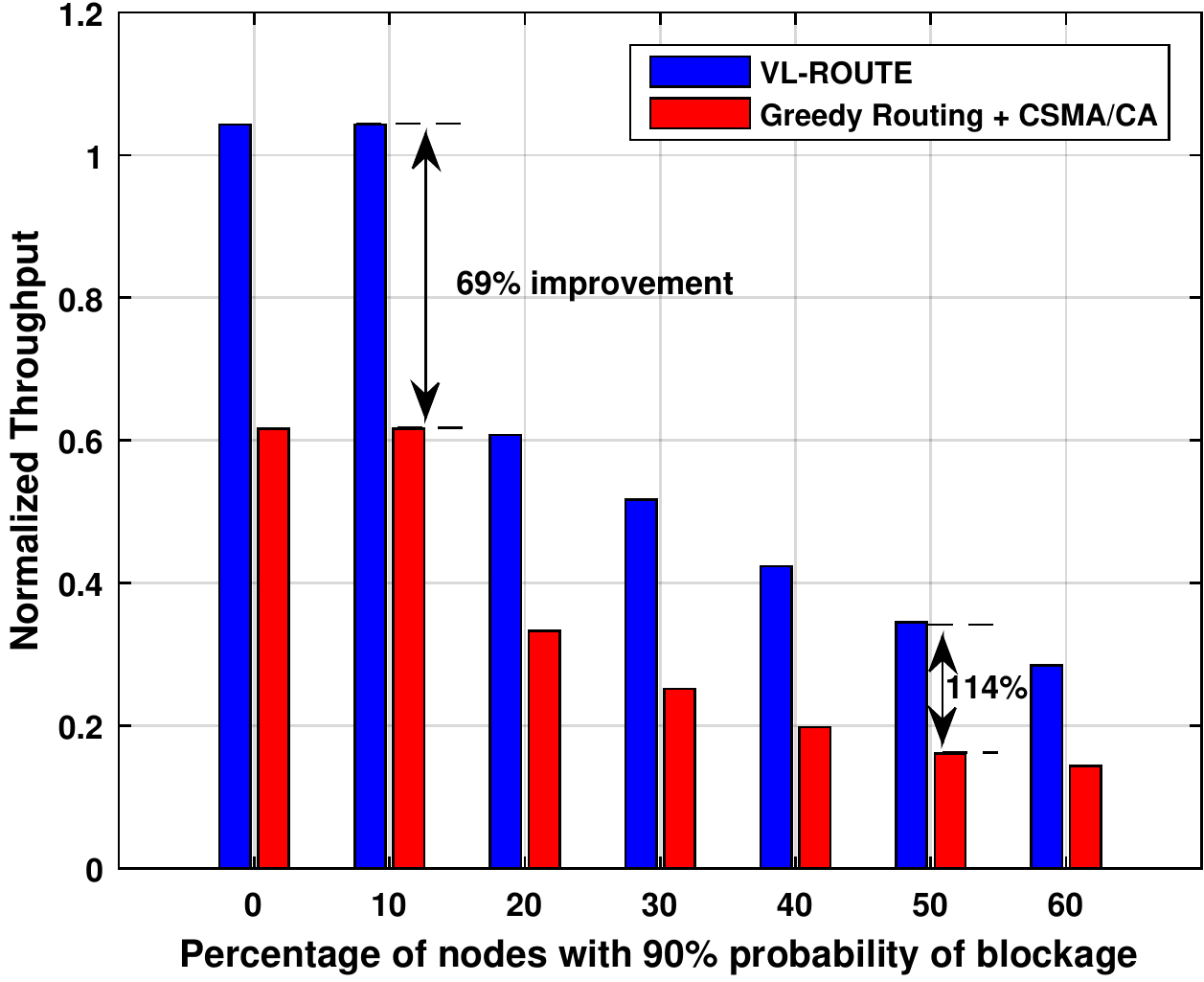}\vspace{ 0 cm}
\caption{Throughput vs Link Blockage}\label{fig:Block} 
\end{figure}

\begin{figure}[h]
\centering
\includegraphics[width=3.3 in]{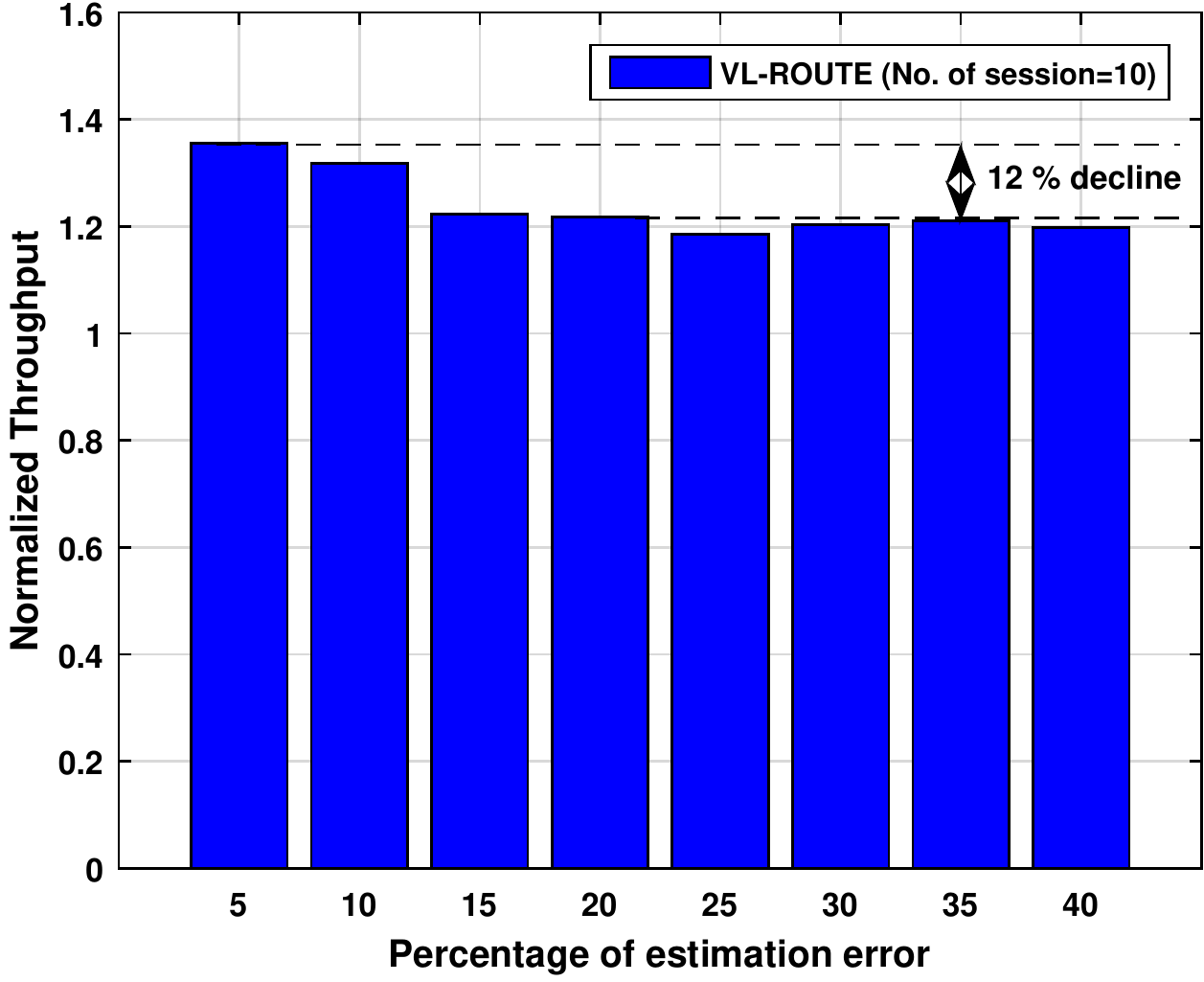}\vspace{ 0 cm}
\caption{Throughput vs Estimation error}\label{fig:Error}  
\end{figure}

\textbf{Estimation Error Analysis.} Since the algorithm depends on the accuracy of estimation of $p_e$ and $p_b$, we evaluate how the VL-ROUTE will be affected with the change in estimation error. It is important to analyze this effect to provide clarity to the reader on how the implementation accuracy affects the performance. Accordingly, we vary the estimation error of both $p_e$ and $p_b$ from $5\%$ to $40\%$ and evaluate the performance by setting the number of sessions to $10$. It can be seen from Fig. \ref{fig:Error} that there is an obvious decrease in performance with the increasing error. The degradation in performance is not drastic (about $12\%$) and the resulting normalized throughput is comparable to one achieved by VL-MAC (see Fig. \ref{fig:Th_grid}). Therefore, having a highly accurate estimation mechanism is advantageous but some error (up to $10\%$) might be acceptable which makes VL-ROUTE a feasible choice for actual deployment.

\section{Conclusion}

To enable \ac{LANET} for several indoor and outdoor applications, there is a significant effort required at the network layer. In this work, we have tackled this problem and proposed a cross-layer routing protocol designed specifically to mitigate challenges like blockage, deafness, and hidden node and to ensure the inherent full-duplex capability of \ac{VLC} is completely utilized. We first recognize that compared to any other ad hoc network, route reliability may be the most important consideration in \acp{LANET} due to the dynamic nature of the \ac{VLC} links. Accordingly, we formulate \ac{RRS} that enables each node in the network to estimate the reliability of route through a neighbor for a given sink. This score is then embedded in a well designed opportunistic \ac{MAC} protocol to exploit the interaction between the network and data-link layer. The measure of reliability along with the cross-layer opportunistic link establishing mechanism provides up to $124\%$ improvement in throughput over GR-CSMA. The effectiveness of the formulated \ac{RRS} is evident when VL-ROUTE outperforms VL-MAC with geographic routing by $21\%$. Additionally, there is an improvement also in the percentage of full-duplex links and the number of packets delivered to the sink.

\bibliographystyle{IEEEtran}
\bibliography{LANET_JJ}

\end{document}